# Towards a Manifesto for Cyber Humanities: Paradigms, Ethics, and Prospects

Giovanni Adorni[*]    Emanuele Bellini[†]

August 3, 2025

## Abstract

The accelerated evolution of digital infrastructures and algorithmic systems is reshaping how the humanities engage with knowledge and culture. Rooted in the traditions of Digital Humanities and Digital Humanism, the concept of *Cyber Humanities* proposes a critical reconfiguration of humanistic inquiry for the post-digital era. This Manifesto introduces a flexible framework that integrates ethical design, sustainable digital practices, and participatory knowledge systems grounded in human-centered approaches.

By means of a Decalogue of foundational principles, the Manifesto invites the scientific community to critically examine and reimagine the algorithmic infrastructures that influence culture, creativity, and collective memory. Rather than being a simple extension of existing practices, *Cyber Humanities* should be understood as a foundational paradigm for humanistic inquiry in a computationally mediated world.

## 1 Introduction

The ongoing digital transformation necessitates a reevaluation of the role of the humanities in a world increasingly shaped by computational technologies. It's not just about digitizing artifacts or using new tools. We need a deeper dialog to redefine how we produce knowledge, preserve cultural memory and engage in civic life.

[*]DIBRIS, University of Genova. Email: `giovanni.adorni@unige.it`
[†]Dept. of Humanities Studies, Roma Tre University. Email: `emanuele.bellini@uniroma3.it`

The present paper introduces and contextualizes the emergence of a novel field of inquiry: the *Cyber Humanities*. Distinct from Digital Humanities, this field addresses the epistemic, ethical and political implications of Artificial Intelligence (AI), algorithmic governance and decentralized information infrastructures. The objective is not merely to integrate technology into the humanities, but rather to critically interrogate the infrastructures that shape meaning and social organization in the algorithmic age.

The debate originates from Digital Humanities, which initially focused on computational tools for cultural analysis, preservation, and dissemination [1, 2]. The *Digital Humanities Manifesto 2.0* [3] emphasized the potential of digital media to transform scholarly communication and research infrastructures, thereby establishing digital humanities as an interdisciplinary practice that transcends traditional print culture. The *Vienna Manifesto on Digital Humanism* further emphasized ethical concerns—such as algorithmic bias, data monopolies, and surveillance—and promoted human-centered technologies aligned with democratic values [4].

The evolution of this landscape—driven by Big Data, AI, Extended Reality (XR), the Internet of Everything (IoE), and distributed ledger technologies—marks the domain of Cyber Humanities. These are not neutral tools: the field investigates their ontological and epistemological implications.

While this emerging field draws on computational paradigms, it also recognizes the continued relevance of interpretive, hermeneutic, and aesthetic traditions. The objective is not to substitute humanistic methodologies, but rather to position them in a dialogic relationship with algorithmic systems and epistemic innovation.

This dialectical engagement is pivotal in ensuring that the richness of humanistic inquiry is not diminished, but expanded through critical examination of technological mediation. This perspective aligns with recent frameworks, such as the Italian CINI Cyber Humanities plan [5] and UNESCO's 2024 report on Artificial Intelligence and cultural heritage [6], which stress algorithmic reflexivity, inclusive innovation, and sustainable governance [7].

Cyber Humanities is not a replacement for Digital Humanities, but a deepening of its critical orientation. It places new epistemologies and competencies at the forefront, which are necessary in a post-digital, post-disciplinary world. At the core of this paper is a *Manifesto for the Cyber Humanities*, presented as a set of principles to guide research, education, and cultural innovation.

The discussion is grounded in three core tenets: (i) algorithmic reflexivity integrated with ethics-by-design; (ii) ecological sustainability as a normative imperative; and (iii) decentralized, open knowledge ecosystems. These form the ethical and conceptual foundation of the Manifesto.

The following sections expand on this framework. Section 2 defines the scope of Cyber Humanities and its technological enablers. Section 3 explores epistemological shifts. Section 4 addresses ethics and AI governance. Section 5 focuses on essential competencies. Section 6 presents case studies. Section 7 outlines the Cyber Humanities Manifesto principles, accompanied by commentary and reflections. Section 8 concludes with a call to action for scholars and institutions.

TABLE I: Comparison between Digital Humanities and Cyber Humanities

| Aspect | Digital Humanities | |
|---|---|---|
| Main Goal | Augment humanities through digital tools | |
| Focus | Digitization, text analysis, tool development | |
| Epistemological Impact | Moderate: method enhancement | |
| Relationship with AI | Instrumental support | |
| Ethical Dimension | Often implicit or external | |
| Governance and Control | Institution-centered digital curation | |

## 2 Cyber Humanities: Definitions and Boundaries

Cyber Humanities is an emerging field at the intersection of humanistic inquiry and computational systems. Its aim is to bridge the "two cultures" [8] and address challenges posed by digital transformation and hybrid threats. Hybrid threats are multifaceted, involving cyberattacks, disinformation, and political manipulation across digital and physical domains. They blur boundaries between war and peace, state and non-state actors, and truth and fabrication. Addressing such threats requires transdisciplinary approaches integrating technological, ethical, cultural, and historical insights [7].

Digital Humanities extend traditional humanities by supporting cultural institutions in digitizing artefacts, analysing texts, and developing tools for scholarly access and dissemination. Its core lies in integrating digital technologies into scholarship. In contrast, Cyber Humanities propose a radical rethinking of the epistemologies, methodologies, and ontologies underlying the humanities.

Cyber Humanities does not simply integrate digital tools into humanistic inquiry but rethinks its epistemological foundations through engagement with computational paradigms—while still valuing traditional interpretive methodologies. This engagement, however, carries risks. To avoid epistemic flattening, technocratic reductionism, or the marginalization of interpretive nuance, computational approaches must be critically assessed.

The Cyber Humanities conceptualizes the humanities domain as a Cultural Cyber-Critical Ecosystem [7], where cultural assets interact with infrastructures and algorithms; recognises AI as a co-constructor of meaning [9]; and embraces hybrid cognitive environments where human and machine intelligences are entangled [10].

As illustrated in Table I, a comprehensive overview is provided of the distinguishing characteristics between Cyber Humanities and Digital Humanities across six key dimensions, encompassing epistemology, ethics, and institutional governance.

## 2.1 Interdisciplinary and Transdisciplinary Approaches

The Cyber Humanities are fundamentally transdisciplinary. Unlike interdisciplinary approaches that merely combine methods from different fields, transdisciplinarity transcends disciplinary boundaries to create new epistemic configurations [11, 12, 13]. This shift involves not only methodological integration but also the redefinition of core questions, vocabularies, and values.

In this context, Cyber Humanities establish a convergence between humanistic inquiry and domains such as computing, cognitive science, engineering, and the social and political sciences [14, 7]. Rather than layering perspectives, this convergence fosters new modes of thought—where ethics and creativity inform code, and algorithms shape theoretical understanding.

This paradigm integrates critical, creative, and technical literacies. Researchers are called to design, analyze, and ethically reimagine algorithmic systems, while adapting conceptual frameworks in response to evolving socio-technical contexts.

In contrast to rigid disciplinary adherence, Cyber Humanities adopt a problem-driven methodology. Knowledge is viewed as dynamic—emerging through iterative inquiry, interdisciplinary collaboration, and socially responsible engagement.

## 2.2 Enabling Technologies

Technological ecosystems in the Cyber Humanities serve two distinct roles: some reshape the epistemological foundations of knowledge—redefining concepts of meaning, interpretation, and authorship—while others function as infrastructural enablers that support and extend these practices. Distinguishing between these dimensions clarifies their respective contributions.

**Epistemologically transformative technologies** include:

**Artificial Intelligence (AI):** Machine learning and GenAI models are not just tools; they function as epistemic agents shaping how knowledge is curated, interpreted, and co-constructed. In Cyber Humanities, AI is both a critical object of study and an active participant in meaning-making [14].

**Big Data:** The growing volume and diversity of data are reshaping how scholars address evidence, temporality, and complexity. Methods like pattern recognition, network analysis, and computational ethnography reveal new dimensions of humanistic inquiry [2, 15].

**Extended Reality (XR) (including Virtual Reality, Augmented Reality, Mixed Reality):** Virtual, augmented, and mixed reality technologies enable immersive reconstructions of cultural memory and participatory storytelling, fostering new forms of epistemic engagement, historiography, and speculative design [1, 14].

**Infrastructural technologies** include:

**Internet of Everything (IoE):** The Internet of Everything (IoE), encompassing variants such as IoT, Underwater IoT, and the Internet of Money, connects objects, spaces, and sensors to enable real-time interaction with cultural environments. Its primary role is to enhance the cyber-physical interface of cultural ecosystems, rather than directly influencing interpretative paradigms [16, 17].

**Distributed Ledger Technologies (DLTs):** DLTs offer new frameworks for authen-

ticating, preserving, and distributing trustworthy cultural artifacts. Though not directly epistemic, they enable novel governance models and promote ethical stewardship of digital heritage [14, 15, 18].

**Cloud Computing:** Cloud systems form the backbone of digital infrastructure, providing the scalability and resilience needed to store, manage, and share cultural data. While foundational to Cyber Humanities practices, their role is primarily operational—supporting digital libraries, preservation systems, and archival platforms.

Consortia such as DARIAH [19] and CLARIN [20] exemplify the importance of infrastructural coordination in ensuring the robustness and sustainability of the cultural research ecosystem in Europe [21].

Having clarified the conceptual and technical landscape of Cyber Humanities, the subsequent section examines how these components converge to reconfigure the epistemological premises of humanistic inquiry.

## 3 The Epistemology of the Cyber Humanities

Cyber Humanities not only introduces new tools into humanistic research, but also prompts a rethinking of knowledge itself in the computational age. The convergence of social, digital, and physical systems—alongside pervasive AI and distributed networks—challenges traditional notions of evidence, interpretation, and authority. This section examines how such dynamics reshape humanistic epistemology, positioning Cyber Humanities as a critical, reflexive science.

### 3.1 Redefining Knowledge in the Digital Era

While philosophical traditions such as Popper's critical rationalism [22] have long emphasized the provisional nature of knowledge, the digital era adds new layers of complexity. Knowledge is no longer seen as a stable archive, but as a dynamic, networked, and context-dependent flow shaped by algorithmic modulation and collaborative infrastructures. This paradigm shift involves several key transformations [23][24]:

- Datafication: The transformation of cultural practices into computable data—quantifiable, analysable, and actionable—reshapes how meaning is structured and accessed;

- Algorithmic mediation: AI systems curate, filter, and generate content, increasingly influencing the creation and circulation of knowledge.

- Distributed authorship: Enabled by technologies like blockchain, this concept challenges traditional notions of authority and ownership in cultural production.

Within the Cyber Humanities paradigm, knowledge emerges from the interplay of humans, AI systems, and digital infrastructures. This calls for a critical epistemology attentive to algorithmic opacity, systemic bias, and shifting architectures of authority. These concerns resonate with Floridi's "infosphere" concept [25], where identity, knowledge, and action are shaped by pervasive information systems.

## 3.2 Computational Thinking and Humanistic Culture

Computational thinking has long been associated with computer science education. However, it is increasingly recognized as a foundational literacy across all knowledge domains [15, 23, 26]. In the Cyber Humanities, it is reinterpreted through a critical and cultural lens.

In the context of the Cyber Humanities, computational thinking is not simply transferred from computer science; rather, it is reshaped — and perhaps even reimagined — to engage with the nuances of humanistic enquiry. Techniques like problem decomposition and abstraction still exist, but are used differently, often to deal with complex, specific questions in literature, philosophy and cultural theory. Algorithms are used not just to process data, but to explore patterns in cultural production. This is evident in networked storytelling, speculative societies and imagined traditions. Yet these same algorithms, which are often neutral, are increasingly scrutinised as cultural artifacts shaped by assumptions, power dynamics and ideological blind spots [14, 23].

The integration of computational thinking with traditional interpretative, hermeneutic, and critical approaches does not constitute a replacement for humanistic methodologies. Rather, it is a synthesis that fosters a novel paradigm of computational digital humanism characterized by analytical rigour coupled with a profound commitment to ethical principles.

## 3.3 Cyber Humanities as a Critical Science

The Cyber Humanities is emerging as a critical science that raises complex and often neglected questions that are becoming increasingly urgent in today's data-driven society. A proactive stance constitutes an essential element of this approach, encompassing the examination of the epistemic shifts driven by algorithmic mediation, the data economy behind platform capitalism, and the ethical dilemmas posed by AI-driven cultural production [27][9].

Rather than treating computational tools as neutral, scholars in this field adopt a reflective perspective. They engage in algorithmic critique, platform analysis and critical design, positioning themselves as both analysts of and co-creators within the socio-technical systems that shape culture and civic life [24][26].

This critical lens raises key questions: How do AI systems reshape what we remember — or forget? Who controls and benefits from the governance of cultural data? And how can cyber-humanistic practices foster equity and ethics rather than reinforce existing asymmetries?

These issues are urgent. They concern justice, agency and responsibility in a world of opaque infrastructures. The Cyber Humanities offer a new kind of science that engages with technology and seeks to shape it in socially meaningful ways.

# 4 Ethics, Rights, and Responsibilities

Integrating computational infrastructures into humanistic inquiry compels a rethinking of ethical responsibility. Positioned at the intersection of culture, technology, and politics, the Cyber Humanities must engage both the potential and the socio-technical risks of digital systems. This section examines the evolving role of digital ethics and introduces cyber citizenship as a civic and ethical paradigm for the algorithmic age.

## 4.1 The Role of Digital Ethics

Ethics is not peripheral in Cyber Humanities; it is foundational, informing all stages of research and practice [9, 28]. Here, digital ethics is both philosophical and operational, addressing how technologies affect cognitive autonomy, epistemic diversity, and ecological balance.

Several ethical imperatives are central to this field: mitigating algorithmic bias and harm [29]; defending user agency against manipulative design [30]; preserving pluralism in AI-generated cultural content [31]; and promoting sustainability in digital infrastructures [28].

This ethical commitment must inform every phase of a project, from data acquisition and algorithm design to dissemination and long-term preservation. This approach aligns with the principles of ethics-by-design [23] and is gaining support from regulatory frameworks, such as the EU AI Act [32] and UNESCO's Recommendation on the Ethics of Artificial Intelligence [27].

Cyber Humanities view ethics not as a constraint, but as a catalyst for innovation and creativity. This perspective challenges researchers to ask not only what can be done, but what should be done—and in whose interest.

## 4.2 Cyber Humanities and Digital Citizenship

In a networked world, the notion of citizenship requires more than technical literacy. In the domain of Cyber Humanities, the concept of "cyber citizenship" has been employed to extend the notion of digital participation, thereby encompassing it as an ethical and political praxis [24, 33]. This encompasses more than merely navigating digital platforms; it necessitates active, critical and responsible engagement with the infrastructures that mediate cultural and civic life.

The concept of cyber citizenship is multi-faceted. The ability to understand and critique the systems that shape cultural production is essential. It has been demonstrated that the platform fosters participation in open knowledge initiatives and the co-creation of digital commons. The approach advocated is founded on a set of values that are rooted in human rights, diversity, and ecological sustainability.

In this context, it is important to note that Cyber Humanities projects should not be regarded exclusively as academic exercises. These interventions can be considered as a form of civic engagement. By encouraging inclusive, participatory spaces, they promote algorithmic accountability and digital justice.

In order to participate in such spaces in a meaningful way, individuals must cultivate a blend of critical and creative competencies. These include algorithmic literacy, ethical awareness, collaborative practice, and the capacity to engage reflectively with AI-mediated narratives. In summary, it is imperative that cyber citizens possess a level of expertise in ethics and culture that is commensurate with their proficiency in code.

## 5 Competencies for Cyber Humanities

The evolution of the field of Cyber Humanities necessitates a rethinking of the competencies required to engage with this emerging discipline. In order to achieve this, scholars and prac-titioners must integrate technical proficiency with critical, ethical and creative thinking, in addition to conventional digital literacy. In a landscape shaped by artificial intelligence (AI), algorithmic mediation and decentralized infrastructures, the ability to navigate, interrogate and reconfigure computational environments is as vital as interpretive and communicative fluency. This section outlines the core competencies needed in cyber-humanities: digital and AI skills, critical thinking and lifelong learning.

## 5.1 Digital Competencies

Recent educational and policy frameworks have emphasized that digital skills are not merely technical [34, 24, 35, 36, 37]. These competencies are foundational to cultural participation, critical reflection and ethical innovation. Digital competencies in the Cyber Humanities comprise a rich and evolving set of skills. These include the ability to locate, interpret and ethically use digital data; to collaborate effectively in AI-mediated environments; and to create, remix and critique digital cultural artifacts. Of particular importance is problem-solving, understood as the capacity to address complex cultural and technological challenges through creative, critical, and interdisciplinary approaches. Furthermore, it is imperative to emphasize the ongoing significance of cybersecurity awareness and ethical responsibility in ensuring the trust and integrity that are fundamental to digital research and cultural practices. This shift in emphasis signifies a transition from a focus on technical proficiency to a more comprehensive digital literacy that is reflexive, value-driven and socially engaged.

It is important to note that these competencies should not be regarded as static checklists. These technologies are evolving in response to emerging developments in related fields, including XR, blockchain and generative AI. In this particular context, the term 'competence' is understood to encompass adaptability, interpretation and accountability.

Recent literature has highlighted an increasing emphasis on the necessity of interaction with intelligent systems being informed, reflective and socially grounded. This notion is emphasized in a growing number of frameworks, particularly those focused on AI literacy [36, 37]. For instance, scholars are required to pose the following question: How are narratives shaped by AI-generated content? The question of ownership and governance of digital cultural memory is a complex one. It is imperative to ascertain the means by which we can guarantee that AI-driven systems will promote diversity rather than perpetuate prevailing structures.

It is important to note that algorithmic literacy in the Cyber Humanities is not limited to technical proficiency; rather, it is a form of interpretative empowerment that enables individuals to critically interrogate, repurpose, and challenge the cultural and epistemic assumptions embedded in computational systems.

It is also important to note that these are not merely theoretical exercises. These challenges are of a live nature and define the future of cultural and scholarly practice. The Cyber Humanities approach to digital competence is not merely concerned with the utilization of technology; rather, it encompasses the critical and responsible shaping of it.

## 5.2 Soft Skills and Critical Thinking

In the event that technical fluency is deemed a prerequisite, it is the possession of critical soft skills that imbues Cyber Humanities with its distinctive character [28]. Analytical reasoning, creative experimentation, intercultural awareness and ethical sensitivity are all essential components of this process.

These competencies enable scholars to challenge the authority of algorithms, deconstruct digital narratives, and co-create new forms of knowledge in hybrid environments. For instance, the ability to identify and resist epistemic closure – where algorithms narrow one's exposure to diverse perspectives – is as crucial as the ability to code itself.

Creativity, meanwhile, is not merely an aesthetic value but a strategic capacity: it enables scholars to imagine alternative modes of cultural production and civic engagement. In the context of a globalised and networked cultural ecosystem, intercultural competence assumes a pivotal role. In this environment, interpretation must take into account a multitude of traditions, narratives and values.

Soft and critical skills are vital in navigating the complex sociotechnical systems of the Cyber Humanities. Analytics, intercultural competence, ethics and creativity enable scholars to move beyond technical functionality to responsible reflection. Critical thinking is especially crucial: helping identify biases, challenge opaque infrastructures and resist manipulative narratives that threaten diversity and civic engagement. These are widely recognized in design justice scholarship [33].

## 5.3 Lifelong Learning and Continuous Training

In light of the rapid advancements in technology, lifelong learning has become imperative rather than a luxury within the domain of the Cyber Humanities. The concept of continuous professional development encompasses a variety of factors, including meta-learning (the ability to acquire and apply learning strategies), reskilling in response to technological shifts, and a robust capacity for autonomous learning.

This learning is increasingly taking place through MOOCs, decentralized platforms, and peer-to-peer communities. These are contexts that value openness, collaboration and agility over rigid institutional pathways.

It is evident that both UNESCO and the European Commission have advocated for the implementation of education systems that are adaptive and supported by AI. These systems have been designed to foster innovation and resilience in the educational sector [38, 39]. In this paradigm, the Cyber Humanities professional is not merely a consumer of knowledge; rather, they become a designer of ethical, participatory and sustainable learning environments.

# 6 Applications and Experimentation

The Cyber Humanities is not merely theoretical constructs; they manifest in a variety of projects, educational innovations and experimental platforms. This section explores the practical applications of the Cyber Humanities principles, examining flagship projects in museums, archives, education and research. The document also examines how educational systems are adapting to global frameworks, as well as exploring how new environments powered by Generative AI are transforming cultural production and learning.

## 6.1 Projects Across Culture and Education

A growing number of flagship initiatives are demonstrating the application of Cyber Humanities, with each initiative reflecting a different facet of the field's ethical and epistemological commitments.

*Museums and Cultural Heritage:* In the museum and heritage domain, the **Smithsonian's Open Access Initiative** has made millions of digitised artifacts freely available for scholarly and creative reuse. The project's use of AI-driven curation systems enables new forms of participatory exploration and cultural reinterpretation, offering a model of decentralized access and engagement [40]. In a similar manner, the **Europeana XR Project** employs immersive technologies to reconstruct cultural heritage sites, thereby encouraging experiential and interactive modes of historical storytelling [41].

*Archives:* In the field of archival studies, the **AI-Assisted Archives Project** at Stanford University employs machine learning algorithms to facilitate the semantic indexing and interpretative retrieval of historical records. In this context, AI is not merely a tool but rather an interpretive co-agent, thereby giving rise to questions concerning transparency, curatorial bias, and the ethical governance of memory [42].

*Education and Research:* In the domain of education, **UNESCO's AI Literacy for Cultural Preservation Programme** is a notable initiative that aims to empower learners with the capacity to utilise AI in a critical and creative manner for the safeguarding of intangible cultural heritage. The promotion of ethical reflexivity and interdisciplinary fluency is fundamental to the ethos of Cyber Humanities education, as articulated in the 2024 UNESCO guidelines [43].

The commonality between these projects is their aptitude for translating intricate sociotechnical challenges into innovative cultural and civic practices. Furthermore, the necessity for collaboration between researchers, technologists, curators and educators is emphasized, thus reinforcing the role of Cyber Humanities as a shared, cross-sectoral endeavor.

## 6.2 Educational Systems and International Frameworks

In response to the demands of a computational society, numerous international initiatives are currently engaged in the active redesign of education systems on a global scale. A number of pivotal frameworks are currently serving as the primary guides for national curricula and vocational training.

Among the most influential frameworks currently shaping digital education policies and curricula is **DigComp**, which integrates AI and data literacy into the baseline digital competence models for citizens [24]. A closely related framework is **DigCompEdu**, a tool designed for educators that promotes the critical use of AI tools in teaching and assessment [34]. Concurrently, joint initiatives by the European Commission and the OECD are progressing a dedicated **AI literacy framework for primary and secondary education** (see reference [39, 44]. Another significant contribution is the **Framework for AI-Powered Learning Environments** developed by the National Center on Education and the Economy, which provides strategic guidance for education leaders, drawing from cross-sectoral best practices in AI integration [45]. Finally, forward-looking curriculum initiatives such as the **CC2020 Task Force**, promoted by ACM and IEEE, highlight the importance of competency-based learning, computational literacy, and interdisciplinary problem-solving as foundational components of education in the algorithmic age [15].

These developments are aligned with the European Digital Education Action Plan 2021–2027, which provides a strategic framework to promote digital skills, inclusion, and innovation across all levels of education in the EU [46].

Across these initiatives, three competencies stand out as essential: the ability to think across disciplines (interdisciplinary fluency), to evaluate the societal implications of technology (ethical reflexivity), and to innovate in hybrid environments that blend physical and digital presence (creative adaptability). As demonstrated in the following section, these characteristics are analogous to the principles outlined in the Manifesto (see next section), thereby providing a framework for their integration into educational practice.

### 6.3 GenAI, Innovation, and the Future of Learning

GenAI is rapidly transforming the production, dissemination and reception of knowledge [28, 39]. GenAI systems are now embedded across cultural and educational ecosystems, from adaptive learning platforms to AI tutors and speculative storytelling engines [47].

AI can personalise learning paths, co-author cultural content and support complex STEM and humanities prototyping. XR classrooms allow learners to engage with historical and speculative worlds. Blockchain credentials aim to enhance trust in educational certification.

Yet this wave of innovation also brings new risks. Algorithmic bias in educational content, unequal access to technology, and the cognitive impact of AI-mediated learning must all be addressed. Here, Cyber Humanities can play a crucial role. This requires sustained engagement from stakeholders to ensure innovation does not outpace responsibility. The challenge is twofold: technological and philosophical. The question that must be answered is how to conceptualize and create learning environments that can facilitate the development of meaning, agency and justice.

The subsequent section proffers a principled response to this challenge. The Manifesto of the Cyber Humanities is presented as a foundational framework intended to guide responsible experimentation, foster ethical transformation, and articulate a shared vision for the future of culture and knowledge.

## 7 The Cyber Humanities Manifesto

This Manifesto frames Cyber Humanities as an emergent epistemological and operational field, drawing on the CINI Cyber Humanities Strategic Plan [5].It advocates responsible, human-centered, sustainable digital practices in knowledge creation, cultural preservation and civic life.

The following ten principles offer scholars, educators, designers and policymakers a framework for interdisciplinary engagement.

## 7.1 The Decalogue

1. **Human-centered Computational Epistemologies:** AI systems must support cognitive autonomy, transparency, and ethical responsibility in the production of knowledge, particularly within cultural and academic contexts.

2. **Algorithmic Reflexivity and Critical Engagement:** The cyber-humanities demand constant examination of algorithmic biases, systemic opacities, and epistemic distortions. Algorithmic reflexivity equips scholars to critically examine how AI systems shape knowledge and to exercise their critical agency in challenging or reshaping those processes.

3. **Ethics-by-Design and Responsible Innovation:** Ethics must be embedded throughout research, education and cultural projects, from the initial planning stage through to deployment. This Ethics-by-Design approach, aligned with UNESCO and IEEE frameworks, ensures human rights, fairness and inclusivity are embedded from the start.

4. **Transparency, Explainability, and Accountability:** AI systems in cultural and academic contexts must be transparent, interpretable, and ethical. In Cyber Humanities, explainability means making visible how algorithmic systems influence interpretation and cultural memory. This requires open design practices, critical documentation, and collaboration across disciplines to develop context-sensitive approaches to algorithmic transparency—even where full technical explainability is not yet feasible.

5. **Dynamic, Distributed and Trustworthy Knowledge Ecosystems:** Knowledge must be produced, preserved, and governed in networked, decentralized, participatory systems that promote transparency, accountability, and shared responsibility. Cyber-humanities promote distributed infrastructures as trustworthy epistemological environments for co-creation, cultural stewardship, and long-term preservation.

6. **Digital Sovereignty and Decentralized Cultural Commons:** The cyber-humanities must support decentralized models of cultural stewardship that empower communities to take control of their digital heritage. Digital sovereignty recognizes collective rights over cultural data, archives and memory, ensuring that preservation practices reflect local values and autonomy.

7. **Equity, Diversity, and Inclusion:** The Cyber Humanities must prioritize equitable access for historically marginalised communities and actively promote cultural diversity in digital spaces. Fostering inclusion is essential to counteracting the structural inequalities embedded in global digital platforms.

8. **Transdisciplinary Methodologies:** Research in the Cyber Humanities must cross traditional boundaries to foster hybrid competencies across the humanities, sciences, and creative technologies. Transdisciplinary approaches dismantle epistemic silos and enable collaborative knowledge production across diverse domains.

9. **Trust, Resilience, and Sustainability Awareness:** Cyber-humanities must critically assess the socio-technical ecosystems in which they operate, evaluating their trustworthiness, resilience to disruption, and ecological impact. In order to build eco-responsible infrastructures, we must acknowledge the hidden environmental costs of digital technologies, including cloud services and AI systems.

10. **Lifelong Learning, Meta-Literacy, and Adaptive Creativity:** Cyber-humanities professionals must develop the skills of continuous learning, algorithmic literacy and creative adaptability needed to navigate rapidly evolving digital environments. Lifelong learning and meta-literacy foster resilience and critical engagement, enabling individuals to evolve alongside post-digital knowledge ecosystems.

## 7.2 Open Reflections and Future Prospects

This Manifesto is intended to function as a dynamic and evolving framework, rather than a fixed doctrinal text. The following pathways are to be explored in future:

- The development of ethical AI guidelines, specifically tailored for humanities-based AI applications, is imperative;

- The establishment of decentralized cultural commons is to be achieved by means of the use of Distributed Ledger Technologies, with the objective of democratizing heritage access;

- The promotion of transdisciplinary research infrastructures is to be encouraged, with a view to establishing links between cognitive science, critical data studies and heritage technologies;

- The creation of global Cyber Humanities alliances to encourage inclusive, global perspectives;

- The integration of trust, resilience and sustainability principles is to be implemented across all digital humanities projects.

To implement these principles, specific actions are needed in education, culture, and policy. In education, AI literacy and critical digital ethics should be included in curricula, as recommended by DigComp and AI-Powered Learning Environments. Cultural institutions, such as museums, libraries, and archives, need to prioritize accessibility, transparency, and inclusion in their digital infrastructures. At a policy level, funding and standards must support ethical innovation goals, encouraging open data governance, community-led digital heritage management, and environmentally responsible technologies.

Such implementation requires collaboration among diverse stakeholders: universities and research centres for methodological innovation; cultural institutions for applied experimentation; technical communities for ethical system design; policy-makers for regulatory alignment; and civil society for inclusive governance and public accountability. Nevertheless, the Manifesto does not disregard the potential risks involved. Algorithmic opacity, extractive data practices, and technocentric epistemologies pose significant challenges. The vocabulary of posthumanist and critical design research proffers pertinent instruments for the reimagination of subjectivity, agency, and ethics in this context [48].

The realization of the vision of Cyber Humanities necessitates collaborative engagement at institutional and policy-level. The final section of the text offers a reflection on the broader implications of the research and calls upon the scientific and cultural communities to take action.

# 8 Conclusions

This field goes beyond the Digital Humanities augmentation paradigm, proposing a reconfiguration of epistemological frameworks, ethical imperatives and educational practices.

These principles not only respond to the opportunities presented by AI, augmented reality, big data, and blockchain technologies, but also to the challenges posed by these technologies, including algorithmic opacity, data colonialism, environmental degradation, and epistemic biases.

The present paper has outlined the theoretical foundations and practical imperatives of the Cyber Humanities, a critical field emerging at the intersection of humanistic inquiry and computational systems.

The Cyber Humanities paradigm demands a reconfiguration of epistemology, ethics, and education, transcending the logic of digital augmentation. The proposed Manifesto sets out ten guiding principles, which are grounded in the following: algorithmic reflexivity, environmental responsibility, decentralized knowledge infrastructures, and lifelong learning.

These principles address the opportunities and challenges posed by the algorithmic age, encompassing subjects such as AI, XR, data colonialism, systemic bias, and ecological impact.

The necessity for collective action is paramount in order to address these issues. In order to operationalise this vision, it is essential that scholars, technologists, educators, policymakers, and cultural institutions collaborate, through the implementation of shared infrastructures, ethical design, and inclusive governance.

This reorientation is not merely of an academic nature. It is imperative to recognize the significance of this initiative in the context of civic engagement, cultural preservation, and ecological sustainability. The Cyber Humanities proffer a framework within which to rethink the creation, governance and preservation of knowledge in computational societies, and to do so in a manner that is both responsible and reflexive, and collaborative.